%% file: main.tex
\documentclass[twocolumn]{aastex631}
\usepackage{graphicx}
\usepackage{gensymb}
\usepackage{amsmath}

\newcommand*{\Lw}{\textit{Lightweaver}}

\newcommand*{\Caii}[1]{Ca~\textsc{ii}#1}
\newcommand*{\Mgii}[1]{Mg~\textsc{ii}#1}
\newcommand*{\halpha}{H$\alpha$}

\shorttitle{2.5D Prominence Bright Rims}
\shortauthors{Jenkins et al.}
\graphicspath{{./}{figures/}}

\input{acronyms}

\begin{document}

\title{The Bright Rim Prominences according to 2.5D Radiative Transfer}

\author[0000-0002-8975-812X]{Jack M. Jenkins}
\affiliation{Centre for mathematical Plasma-Astrophysics, Celestijnenlaan 200B, 3001 Leuven, KU Leuven, Belgium}
\email{jack.jenkins@kuleuven.be}

\author[0000-0002-2299-2800]{Christopher M. J. Osborne}
\affiliation{SUPA School of Physics and Astronomy, University of Glasgow, Glasgow G12 8QQ, UK}

\author[0000-0002-1190-0173]{Ye Qiu}
\affiliation{Institute of Science and Technology for Deep Space Exploration, Nanjing University, Suzhou 215163, China}

\author[0000-0003-3544-2733]{Rony Keppens}
\affiliation{Centre for mathematical Plasma-Astrophysics, Celestijnenlaan 200B, 3001 Leuven, KU Leuven, Belgium}

\author[0000-0001-7693-4908]{Chuan Li}
\affiliation{Institute of Science and Technology for Deep Space Exploration, Nanjing University, Suzhou 215163, China}
\affiliation{School of Astronomy and Space Science, Nanjing University, Nanjing 210023, China}
\affiliation{Key Laboratory for Modern Astronomy and Astrophysics (Nanjing University), Ministry of Education, Nanjing 210023, China}



\begin{abstract}

Solar prominences observed close to the limb commonly include a bright feature that, from the perspective of the observer, runs along the interface between itself and the underlying chromosphere. Despite several idealised models being proposed to explain the underlying physics, a more general approach remains outstanding. In this manuscript we demonstrate as a proof-of-concept the first steps in applying the \Lw{} radiative transfer framework's 2.5D extension to a `toy' model prominence + VAL3C chromosphere, inspired by recent 1.5D experiments that demonstrated a significant radiative chromosphere\,--\,prominence interaction. We find the radiative connection to be significant enough to enhance both the electron number density within the chromosphere, as well as its emergent intensity across a range of spectral lines in the vicinity of the filament absorption signature. Inclining the viewing angle from the vertical, we find these enhancements to become increasingly asymmetric and merge with a larger secondary enhancement sourced directly from the prominence underside. In wavelength, the enhancements are then found to be the largest in both magnitude and horizontal extent for the spectral line cores, decreasing into the line wings. Similar behaviour is found within new \textit{Chinese H$\alpha$ Solar Explorer} (CHASE)/\textit{H$\alpha$ Imaging Spectrograph} (HIS) observations, opening the door for subsequent statistical confirmations of the theoretical basis we develop here.

\end{abstract}



\section{Introduction} \label{s:intro}
Solar prominence bright rims have been observed for decades, and in response numerous models have been constructed that aim to explain their origin \citep[][and references therein]{Vial:2015}.
For example, \citet{Heinzel:1995} argued that the underside of a solar prominence scatters more light towards the observer than its overside, making this lower region appear brighter.
\citet{Paletou:1997} then challenged this conclusion, suggesting that such a process would be unable to explain the magnitude of intensity increase measured for prominence bright rims.
Since then, \citet{Panasenco:2010} and others have suggested the enhancement may have a chromospheric origin, being instead related to the viewing angle on underlying fibrillar structures that have different thermodynamic properties.
The exploratory study of \citet{Kostik:1975} also considered the phenomenon to be of chromospheric origin, suggesting that radiation trapping--the bouncing back and forth of radiation between the chromosphere and an overlying prominence--may be responsible.
Our historical inability to construct self-consistent models of solar prominences, inclusive of radiation, has hindered attempts to validate any one of these theories.

The recent study of \citet{Jenkins:2023} highlighted, however, that the classical 1.5D approximation failed for the elevated atmospheres of solar prominences on account of a significant radiative trapping that existed between it and the underlying chromosphere \citep[see also][]{Paletou:1993}.
The cause of this trapping was found to be the 1.5D plane-parallel approximation that implicitly assumes the prominence to be infinitely horizontal in extent.
Hence, all upwards directed radiation released from the chromosphere encounters the prominence, regardless of the angle of the quadrature, and a non-negligible flux of radiation is then scattered back towards the chromosphere.
The result is that this radiation bounces back and forth pumping the populations of the prominence, but also the underlying chromosphere.
This trapping was so significant, that it drove the core of the Hydrogen~\halpha{} line into positive contrast against the illuminating chromosphere.
This is in direct contradiction to the characteristic appearance of \halpha{} filaments as dark channels snaking across the surface of the Sun \citep[\textit{e.g.},][]{Parenti:2014}.

This experiment provided a clear indication that a radiative connection existed between prominences and their underlying chromospheres, at least in 1.5D.
In multi-dimensions, prominences are not infinite in extent but neither are they infinitely thin.
Moreover, the length of a prominence can often be an order of magnitude longer than its width.
Taken together, it is probable that a non-negligible amount of radiation is in any case scattered back towards the chromosphere.
The aim of this manuscript is to therefore explore and quantify this radiative interaction.

We introduce the model geometry and 2.5D radiative transfer considerations in Section~\ref{s:methods}, the principle results are shown in Section~\ref{s:results}, before being summarised and discussed in Section~\ref{s:discussion}.

\section{Methods} \label{s:methods}

\subsection{Model Geometry} \label{ss:modelGeometry}

\begin{figure}
	\centering
	\resizebox{\hsize}{!}{\includegraphics[clip=,trim=0 0 0 0]{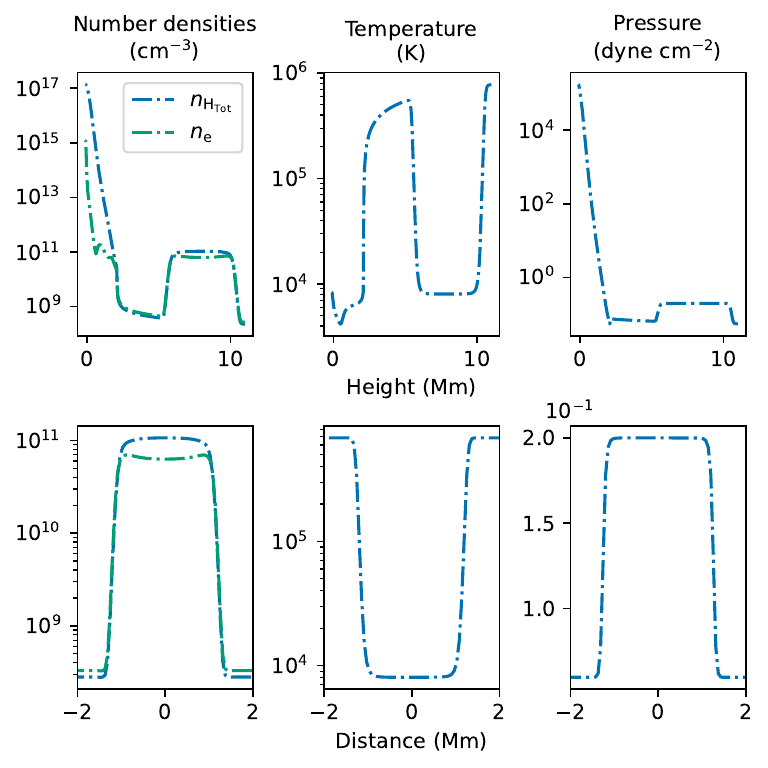}}
	\resizebox{\hsize}{!}{\includegraphics[clip=,trim=0 6 0 0]{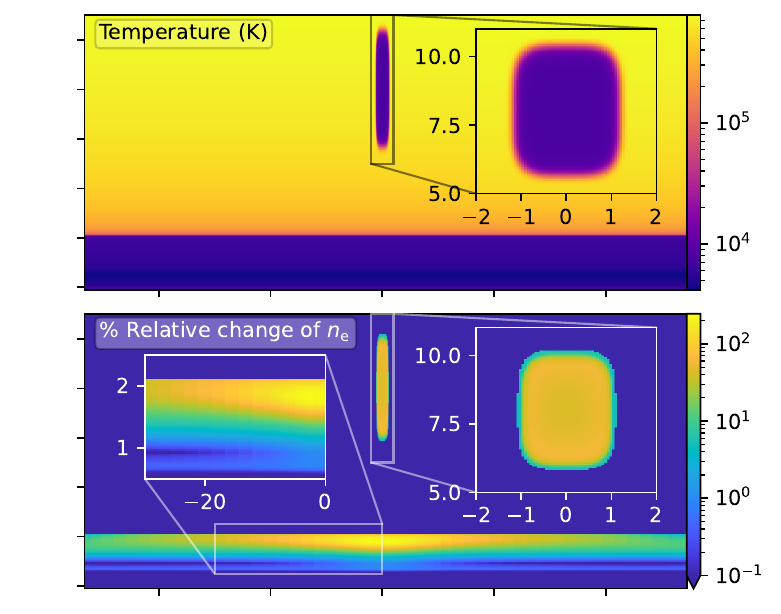}}
	\resizebox{\hsize}{!}{\includegraphics[clip=,trim=0 0 0 3]{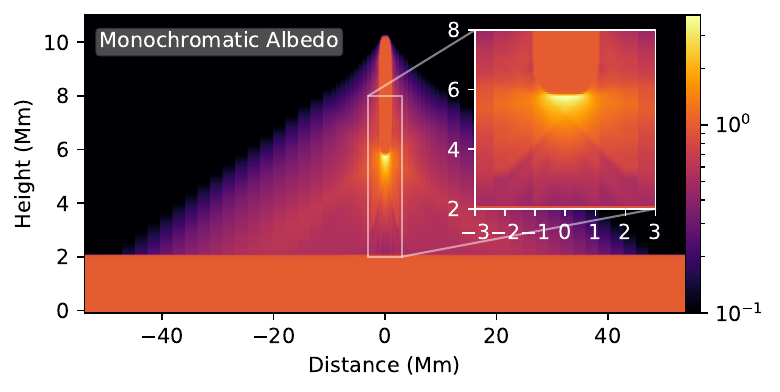}}
	\caption{The stratification of the `toy' prominence model and background atmosphere. The first and second rows present the vertical and horizontal variation, respectively, of the electron and total Hydrogen number densities, temperature, and pressure in the model atmosphere. The bottom three panels present the spatial distribution of temperature, \% relative change of $n_\mathrm{e}$ (photoionisation), and the angle-integrated monochromatic albedo for the head of the Lyman continuum, respectively, with inset panels zooming in on specific details.}
	\label{fig:primitiveStratification}
\end{figure}

We aim herewith to quantify the radiative interaction existing between the solar chromosphere and an isolated solar prominence suspended within the solar corona.
For the chromospheric portion of the model, we adopt a modified VAL3C stratification that extends from 92~km below the photosphere to 11~Mm into the solar corona, encompassing the temperature minimum, chromosphere, and transition region in between \citep[][]{Vernazza:1981}.
Following the approach adopted by RADYN and discussed in \citet{Carlsson:2023}, the hydrostatic extrapolation of the underlying radiative equilibrium assumes a vertical rather than short-loop geometry.
This then invariantly spans a range of -55\,--\,55~Mm in the horizontal direction.

For the prominence atmosphere, we recently extracted self-consistent prominence atmospheres from \texttt{MPI-AMRVAC} \ac{MHD} simulations, coupling them to semi-empirical models of the solar chromosphere \citep[][]{Jenkins:2023,Keppens:2023}.
Here we instead construct a `toy' prominence stratification outside of hydrostatic equilibrium and embed it within the modified VAL3C atmosphere \citep[similar to the approach of][]{Diazbaso:2019}.
In this way we can concentrate the necessary resolution at the \ac{PCTR} interface.
We chose the minimum temperature of the prominence core to be a characteristic 8000~K at 0.2~dyn~cm$^{-2}$, with a geometrical extent of 2~Mm and 5~Mm in the horizontal and vertical directions, respectively, centered on (0,8)~Mm.
We set the shape of the prominence to that of a `squircle', a unit sphere within the $\ell^p$ space, to negate numerical oscillations in the interpolation at the sharp corners of a square,
\begin{equation}
    |x|^p +|y|^p = r^p,
\end{equation}
where $x$ and $y$ are the Cartesian coordinates, $p$ is the order of the space, and $r$ the radius of a chosen unit sphere. The specific $\ell^p$ space of order $p=4$ and radius  $r=1$ is named the `squircle', but any radius will benefit from the continuity characteristic.
This prominence core is then smoothly connected to the solar corona using the hyperbolic tangent function  \citep[cf.][]{Xia:2012},
\begin{align}
	f(x,y) &= \frac{\mathrm{tanh}(s \times (r(x,y) - x_\mathrm{off})) + y_\mathrm{off}}{y_\mathrm{off} + 1},\\
	T(x,y) &= T_\mathrm{core} + (f(x,y) \times (T_\mathrm{VAL3C}(x,y) - T_\mathrm{core})),
\end{align}
where $s=2$ controls the `steepness' of the $\mathrm{tanh()}$ profile, $x_\mathrm{off} = 2.5$~Mm, $y_\mathrm{off} = 1$~Mm, and $T_\mathrm{core} = 8000$~K.

By adopting such a `toy' model, the gradients in the thermodynamic properties that connect the prominence core with the solar corona, the \ac{PCTR}, are hydrodynamically idealised and hence differ from the 1D \ac{MHS} class of (K-S type) thread solutions from \citet{Heinzel:2001} that have been widely adopted \citep[][]{Kippenhahn:1957}.

Such solutions do not consider the same temperature ranges as we do and so, in a more hybrid manner, we first ensure the connecting hyperbolic tangent profile closely traces the \ac{MHS} solutions at the low temperatures of $<20$~kK.
Above this temperature, the hyperbolic tangent is much sharper than the \ac{MHS} models and hence closer to those of \citet{Gouttebroze:2006}, but such temperatures lie outside the range considered applicable for the transitions that we will explore in this manuscript \citep[cf.][]{Carlsson:2012}.

The resulting model domain, alongside 1D cuts of the vertical and horizontal model stratifications, is presented in Figure~\ref{fig:primitiveStratification}.
The model has a vertical resolution of 300 points with a variable $\delta y$ established following an optimisation routine that distributes and focus these points in locations of strong atmospheric gradients\footnote{\href{https://github.com/jaimedelacruz/pTau/tree/main}{The pTau Python package of J. de la Cruz Rodriguez}}.
The horizontal dimension has a resolution of 131, with a $\delta x$ that focuses the highest resolution around the prominence and varies symmetrically about 0~Mm: $\approx57$~km out to 2~Mm, then 0.8~Mm out to 9~Mm, and finally 2.25~Mm out to the edges of the model.

\subsection{Radiative Transfer} \label{ss:radiativeTransfer}
Throughout this work we will primarily make use of the \Lw{} framework.
We have previously applied such a framework to the synthesis of a range of solar prominence atmospheres under the 1.5D plane-parallel approximation \citep[][]{Jenkins:2023}.
Here, we use instead the 2.5D extension recently reported on in \citet{Osborne:2022b}, but without the time-dependent evolution of the rate equations as they employ.
Here, the `.5' indicates that all vectors in the 1D and 2D geometries contain two and three components, respectively, that are considered invariant in this extra dimension.

The latest version of \Lw{} contains domain decomposition in the vertical direction to facilitate deployment on massively-parallelised machine architectures.
This is implemented as sub contexts to the \texttt{Context} object in the Python frontend, such that each subdomain acts as fixed boundary conditions to their adjacents.
This has the consequence that information has a finite travel distance per iteration, slightly lengthening the time taken to reach convergence.
The bottom boundary maintains the diffuse thermalised radiation approximation, and the upper boundary remains `open' - we consider no incoming radiation from the solar corona.

\Lw{} in 2.5D has two formal solvers implemented, linear or BESSER following the implementation of \citet{Stepan:2013}, of which we employ the latter here.
Despite focusing the resolution of the model to the prominence and \ac{PCTR}, we have adopted linear interpolation as initial tests yielded non-conservation of energy across sharp intensity gradients when using BESSER.
This will lead to increased diffusion in the propagation of radiation throughout the model in comparison to a more advanced interpolation method.

Our early tests found that the exact angular quadrature assumed for the discrete propagation of radiation through the domain had direct consequences on the solution.
The low order A-2\,--A-8 quadrature sets of \citet{Carlson:1963} \citep[see also][]{Bruls:1999}, the anisotropy-optimised quadratures of \citet{Stepan:2020} and \citet{JaumeBestard:2021}, and the Gauss-Legendre(-Trapezoidal in multi-D) sets - commonly employed as ground-truth benchmarks - led to the `spotlighting' of the prominence-backscattered light on the underlying chromosphere \citep[][]{Paletou:1995}.
That is, a lack of, asymmetric, or non-uniformly weighted angular resolution, respectively, led to the focusing of the significant backscattered flux into discrete locations at chromospheric heights.
Such issues have thusfar gone unreported on account of solar atmospheric models generally not considering the discrete and isolated plasma condensates at elevation that are prominences.
To overcome this, we implemented the HEALPix equal weight spherical discretisation of \citet{Gorski:2005}, that has proven useful in other radiative transfer contexts \citep[such as][]{DeCeuster:2020}.
The utility here lies in the discretisation not introducing any optimisations to the weighting of the individual rays, and so more evenly distributing all of the aforementioned backscattered radiation.
The explicit demonstration of these limitations will be presented in full in a subsequent study.
We choose the HEALPix set of order 4, containing 100 rays over a half sphere, of which eight contributed halfed weights to the integration to satisfy the symmetry condition.

The bright-rim phenomenon is most commonly reported for prominences close to the limb from the perspective of the observer.
At such a position, limb-darkening plays a non-negligible role in the appearance of the background and should be taken into account \citep[][]{Heinzel:1995}.
At such an extent, limb-darkening is not only angularly dependent but also spatially so; at small $\mu$ angles the curvature of the Sun is non-negligible within the \ac{FOV}.
The model itself assumes a Cartesian geometry, hence the spatial component cannot be taken into account within the model domain.
The illumination of the model according to limb darkening is instead included within the boundary conditions \citep[as in][]{Jenkins:2023}.
Fixing the lateral boundaries in this way ensures the horizontal rays from the HEALPix quadrature, problematic in a periodic model, receive the necessary information.
Furthermore, for very small $\mu$ angles, the solar limb will be present within the \ac{FOV} and a self-consistent transition of the model filament to prominence will be possible.

Exactly as for our previous works we consider the same 5 level + continuum hydrogen atom with 10 bound-bound transitions, 5 level + continuum calcium~\textsc{ii} atom with 5 bound-bound transitions, and 10 level + continuum magnesium~\textsc{ii} atom with 15 bound-bound transitions. Although we include all of these atoms (hereafter H, \Caii{}, \Mgii{}) and their transitions in the convergence, we will only focus on the common H and \Caii{} transitions within this manuscript and reserve the more complex \Mgii{} to a more comprehensive follow-up study. Convergence is considered satisfied when the angle-averaged radiation field $J$, population levels, and \ac{PRD} $\rho$ reach a maximum relative absolute difference between iterations of $5\times10^{-3}$, $1.5\times10^{-3}$, and $1\times10^{-4}$, respectively.

During the iterations of the population levels, we additionally iterate for the unknown $n_\mathrm{e}$ using the \ac{NLTE} prominence tables of \citet{Heinzel:2015} as our initial conditions.
This is an important step to take photoionisation by the radiation illuminating the prominence into consideration and, as we shall show later, the influence of its backscattering on the chromosphere.
As this iteratively alters the pressure of the equilibrium, we modify $N_\mathrm{H_{Tot}}$ to maintain the pressure initial conditions and also update the other population levels accordingly \citep[cf.][]{Heinzel:1995, Paletou:1995}. 

\section{Results} \label{s:results}

\subsection{Converged $n_\mathrm{e}$} \label{ss:convergedNe}
In 1.5D radiative transfer models of solar prominences, iterating the electron number density during convergence leads to an increase in their population at prominence extremities on account of photoionisation by, for example, the Lyman and Balmer continua \citep[cf.][]{Heinzel:2014a, Zapior:2016}.
For the 2.5D model here, Figure~\ref{fig:primitiveStratification} shows that the iterated electron number density at the edges of the prominence is indeed enhanced relative to the core population, but more modestly than reported in the 1.5D models.
Similarly, there appears to be almost no difference in the magnitude of $n_\mathrm{e}$ within the lower, sun-facing \ac{PCTR} versus the space-facing side, with the former being a mere 1\% larger.
Of particular interest is the $n_\mathrm{e}$ enhancement in the upper layers of the underlying chromosphere that decreases in magnitude with increasing horizontal distance from the prominence.
In much the same way, the extent of the enhancement in depth is greatest directly underneath the prominence and decreases with horizontal distance.

In the same Figure, we show the monocromatic albedo ($A_\nu$) at the head (912~\AA) of the photoionising Lyman continuum (LyC) following,
\begin{equation}
	A_\mathrm{\nu} = \frac{\int I^\mathrm{down}_{LyC} \, d\mu}{\int I^\mathrm{up}_{LyC} \, d\mu},
\end{equation}
wherein the underside of the prominence is shown to scatter a significant amount of incident radiation back towards the solar chromosphere. 
It is the back and forth of this ionising radiation that leads to the increased $n_\mathrm{e}$ in both the prominence and its underlying chromosphere.

\newpage
\subsection{Spatial and Angular $I_\mathrm{e}$ variations} \label{ss:spatialIe}

\begin{figure}
	\centering
	\resizebox{\hsize}{!}{\includegraphics[clip=,trim=0 0 0 0]{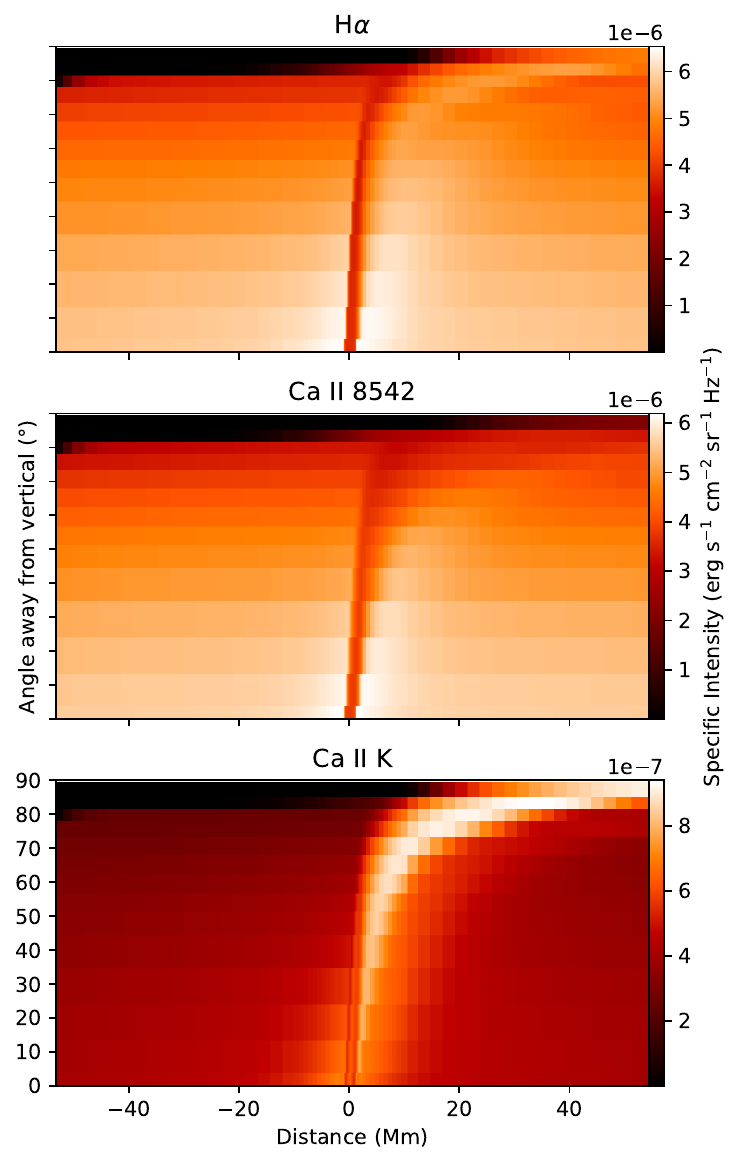}}
	\caption{Spatial and angular variation of the \halpha{}, \Caii{~8542~\&~K} spectral line intensities $I_\mathrm{e}$ emerging from the `toy' prominence model of Figure~\ref{fig:primitiveStratification}. For each spectral line, from the bottom to top-most row the \ac{LOS} rotates from vertical to horizontal. Symmetric brightenings are present either side of the filament absorption for $\mu_\mathrm{z}=1 \rightarrow \theta_\mathrm{z} = 0\degree$, becoming increasingly antisymmetric and comprised of fine structures, that vary depending on the spectral line, as the viewing angle increases.}
	\label{fig:angularTransitions}
\end{figure}

\begin{figure}
	\centering
	\resizebox{\hsize}{!}{\includegraphics[clip=,trim=0 0 0 0]{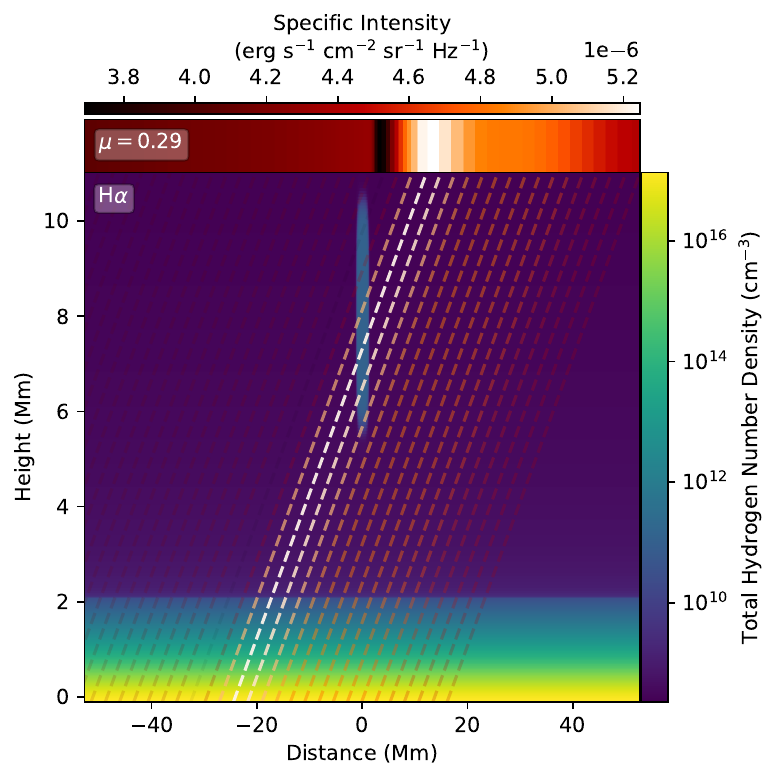}}
	\caption{Source of intensity enhancements present next to central filament absorption.  \textit{Top panel}; Emergent intensity $I_\mathrm{e}$ of the \halpha{} transition at a slightly inclined viewing angle of $\mu = 0.29$ to the vertical. \textit{Bottom panel}; Model total hydrogen number density, with dashed lines overlaid to indicate the intersection between the \ac{LOS} and features within the domain. The lines are additionally coloured according to the intensity within the upper panel, and their transparency set to emphasise the brighter \acp{LOS}. The broader \halpha{} intensity enhancement comes from the chromosphere beneath the prominence, whereas the brighter enhancement comes directly from the prominence underside.}
	\label{fig:halphaBrightOrigin}
\end{figure}

\begin{figure}
	\centering
	\resizebox{\hsize}{!}{\includegraphics[clip=,trim=0 0 0 0]{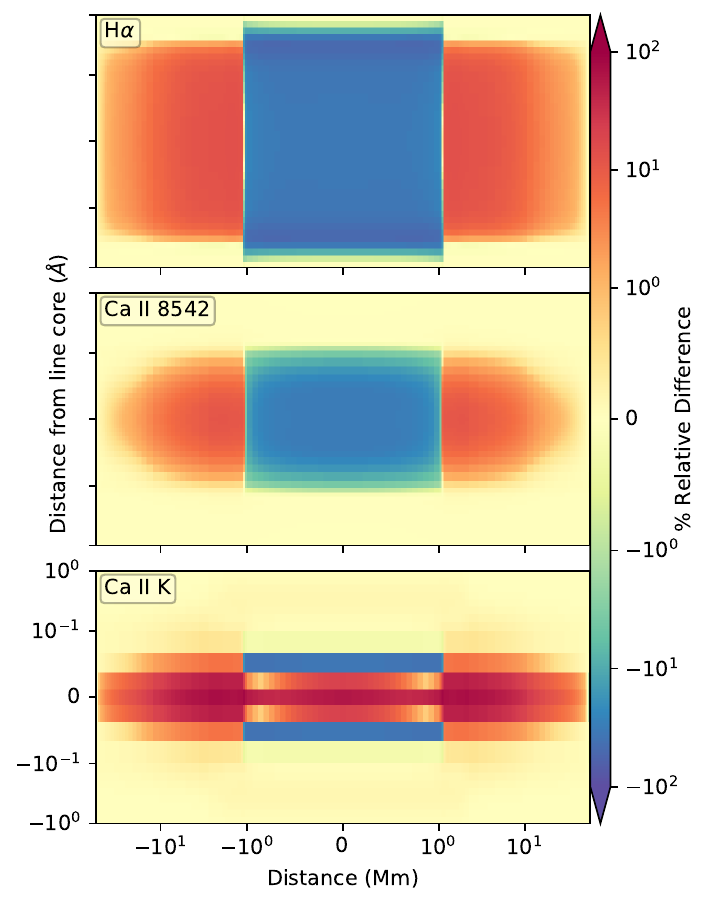}}
	\caption{Spatial and spectral variation of the \halpha{}, \Caii{~8542~\&~K} spectral line \% relative differences in $I_\mathrm{e}$ for $\mu_\mathrm{z} = 1$ compared to a chromosphere-only domain, positive indicating brighter and vice versa.}
	\label{fig:spectralTransitions}
\end{figure}

The top panel of Figure~\ref{fig:angularTransitions} presents the line core emergent intensity $I_\mathrm{e}$ of the \halpha{} transition at a range of observing angles into the top of the `toy' prominence model domain of Figure~\ref{fig:primitiveStratification} \citep[cf. Fig.~4 of][]{Libbrecht:2021}.
The bottom row of this panel describes the intensity emerging exactly vertically ($\mu_\mathrm{z}=1 \rightarrow \theta_\mathrm{z} = 0\degree$).
The absorption profile for the filament at 0~Mm is apparent but made more obvious by the larger intensities (in comparison with the edge of the synthesis domain at 55~Mm) that border it.
The y-axis of the same plot then describes how this emergent intensity changes for an increasing viewing angle of the observer (decrease $\mu_\mathrm{z}$) towards the horizontal.
The intensity increase flanking the absorption profile becomes increasingly asymmetric as the \ac{LOS} is inclined through 0\,--\,40$^{\degree}$, migrating in the + Distance direction.
After 40$^{\degree}$, this enhancement becomes simultaneously weaker and more diffuse as a combined consequence of the inclined viewing angle (limb-darkening) and the coarser spatial resolution through which the radiative transfer is calculated, respectively.
At approximately 60$^{\degree}$ a second bright feature is identifiable now closer to the filament absorption, and becomes increasingly visible against the first as the viewing angle increases past this point.
As the inclination approaches 80$^{\degree}$, the portions of the \ac{LOS} that are at X~$<-50$~Mm begin to miss the solar disk altogether and thus have zero intensity.
As the inclination approaches 90$^{\degree}$, the prominence transitions from the filament to prominence projection, instead appearing bright against the background of space.
The remainder of the Figure presents the equivalent appearance and angular-variation for the \Caii{~8542~\AA~\&~K} line cores.
The broader intensity enhancement remarked to flank the central absorption of the filament is present in each case.
The thinner enhancement that appeared past ~60$\degree$ is entirely absent for the \Caii{~8542~\AA} intensity.
For \Caii{~K}, on the other hand, this feature is present almost from the moment the \ac{LOS} deviates from the exact vertical and strongly dominates the emergent intensity thereafter, essentially rendering the prominence as an emission feature against the background solar chromosphere.
We will discuss this in more detail in Section~\ref{s:discussion}.


In the upper strip of Figure~\ref{fig:halphaBrightOrigin} we zoom in on the single viewing angle of $\mu = 0.29$ for the \halpha{} emergent intensity.
Here, the intensity limits are now tuned to the maximum and minimum of this singular strip rather than all strips, better highlighting the variations herein.
At this viewing angle, the two aforementioned bright components are clearly identified as a superposition of a broad enhancement that spans from the filament absorption to approximately 40~Mm, and a thinner brighter enhancement that spans approximately 5\,--\,15~Mm.
In the bottom panel of total hydrogen number density, from the traced rays, we see that the broader enhancement is sourced from the chromosphere underneath the prominence whereas the brighter enhancement is instead from the lower portions of the prominence body itself.
Furthermore, at this angle, the filament absorption appears to correspond to only the upper portions of the prominence body.

\subsection{Spectral $I_\mathrm{e}$ variations} \label{ss:spectralIe}

In Figure~\ref{fig:angularTransitions} we found a spatially symmetric enhancement flanking the central filament absorption feature at a viewing angle of~0$^{\degree}$.
In Figure~\ref{fig:spectralTransitions}, we expand to include the wavelength dimension and present the \% relative difference in $I_\mathrm{e}$ against a pre-relaxed chromosphere-only domain, where the spatial and spectral scaling is now log\,--\,log to emphasise the spatial and spectral cores of the prominence and spectral lines, respectively.
The model of Figure~\ref{fig:primitiveStratification} contains zero velocities throughout the domain, so the spectral profiles of all the transitions shown are expectedly symmetric about their rest wavelength for all spatial positions.
The intensity variations in both space and wavelength are symmetric about $0$ for $\mu=1$, yet the enhancement is wavelength dependent with the largest spatial extent being in each case at the line core wavelength.
For \halpha{}, we see that as the spatial distance from the prominence body increases, the relative difference drops to 0 closer to the line core, an identical feature yet more pronounced for both the \Caii{~8542~\&~K} lines.
For the \Caii{~K} transition, a more complex situation appears present, with the gradated central line core enhancements giving way to relative decreases at larger $\Delta \lambda$.
The wavelength sampling employed in the spectral integration of the radiative rates is quite insufficient for further analysis of the \Caii{~K} transition, though this will be discussed in more detail in Section~\ref{s:discussion}.

\begin{figure*}
	\centering
	\resizebox{\hsize}{!}{\includegraphics[clip=,trim=0 0 0 0]{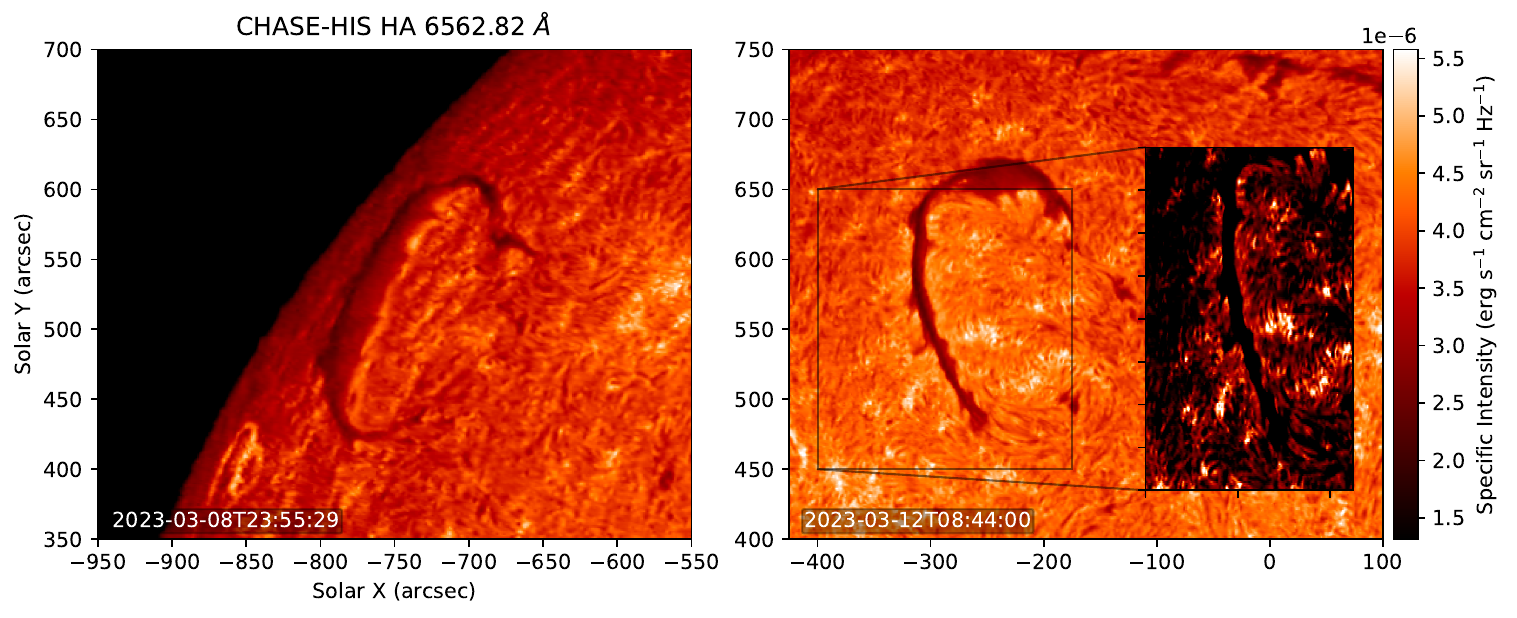}}
	\caption{Two \ac{CHASE}/\ac{HIS} observations of the same prominence exhibiting the bright-rim phenomenon on 2023-03-08 and 2023-03-12. At the earlier time, the filament was located at $\approx~60^{\degree}$. For the later time, the filament was located at $\approx~20^{\degree}$. The intensity of the filament observation in the inset of the right panel has been saturated to emphasise the brightenings on either side of the filament body.}
	\label{fig:chase_brightRim}
\end{figure*}

\section{Summary and Discussions} \label{s:discussion}
It was previously shown by \citet{Jenkins:2023} that a 1.5D approach to a combined chromosphere + prominence atmosphere suffers from artificial radiation trapping/pumping caused by the infinite plane-parallel assumption \citep[with relevance to \textit{e.g.},][]{Diazbaso:2019}.
This indicated that a radiative connection existed between the solar chromosphere + prominence, as previously hinted at by \citet{Paletou:1995}.
To further explore this a 2.5D domain was required so that both the partial radiation trapping and the finite extent of the prominence could be represented within the same model.
The aim of this manuscript was to adopt the 2.5D \Lw{} radiative transfer framework, apply it to a specific case-study `toy' prominence model, quantify the degree of chromosphere\,--\,prominence connection present, and demonstrate the utility of this approach in further exploring the as-yet insufficiently explained `bright-rim' prominence phenomenon.

In relaxing the atomic transitions to statistical equilibrium in 2.5D, aided by the initial conditions available within the \citet{Heinzel:2015} 1D \ac{NLTE} radiative transfer solutions, we began by confirming the electron number density $n_\mathrm{e}$ to be significantly modified within the prominence \citep[as previously stated by][and subsequently shown by \citealt{Leger:2007b}, to be a consequence of photoionisation]{Paletou:1995}.
Unlike the 1D models of \citet{Heinzel:2014a} or \citet{Zapior:2016}, however, the degree of enhancement that both \citet{Leger:2007b} and we find within the prominence and/or \ac{PCTR} appears less pronounced.
That said, the models of \citet{Gouttebroze:1993} already demonstrated that such enhancements can vary significantly even in 1D, and our `toy' model represents but a single idealised case study; the parameter space of photoionisation behaviour in multidimensional prominence models requires further exploration.
As our 2.5D domain integrates both the chromosphere and the prominence, we confirm and expand upon the predicted prominence backscattering of \citet{Paletou:1993}, demonstrated here using the albedo of the Lyman continuum head in Figure~\ref{fig:primitiveStratification}, in its capacity of pumping the population of $n_\mathrm{e}$ within the chromosphere. This excess is most pronounced underneath the prominence and decreases with increasing distance. 

In Figure~\ref{fig:angularTransitions} we found that the locations of increased $n_\mathrm{e}$ within the chromosphere overlap with the regions of increased emergent intensity $I_\mathrm{e}$ that flank the central filament absorption for a viewing angle of $\mu_\mathrm{z} = 1$.
By increasing the viewing angle towards the horizontal, we saw that such brightenings become increasingly asymmetric and skewed in the direction of the observer.
At $\mu_\mathrm{z}=0.29$, and as shown in Figure~\ref{fig:halphaBrightOrigin}, a second more intense brightening becomes clearly visible closer to the filament absorption signature.
In the same figure, we showed that the more diffuse brightening originates from the underlying chromosphere \citep[as previously argued by \citealp{Kostik:1975} and][]{Panasenco:2010} whilst the brighter and more-localised intensity enhancement is sourced in the underside of the prominence itself \citep[closer to the interpretation of][albeit contested by \citealp{Paletou:1997}]{Heinzel:1995}.

In Figure~\ref{fig:spectralTransitions}, we explored the difference in appearance of the diffuse brightenings that flank the filament absorption for a viewing angle of $\mu_\mathrm{z} = 1$.
For each of the spectral lines shown, we found their \% relative difference to range between a few \% to more than 100\%.
These brightenings then have the largest extent for the line core wavelength, likely as a consequence of the prominence being the most optically-thick and/or scattering in the line core.
As such, the exact magnitude and extent of these brightenings must depend on numerous aspects of the prominence atmosphere, and will require a theoretical effort as by \citet{Gouttebroze:1993} to understand them better.


The next step in the process of proving the theoretical aspects highlighted in this manuscript is to compare directly with observations.
We envision a statistical study, that will take into account a range of observed prominences that exhibit the bright-rim phenomenon.
A host of historical observations are available through the GONG \halpha{} network \citep[][]{Harvey:2011}.
Of particular interest is the \acf{HIS} instrument onboard the \acf{CHASE} satellite, a slit spectrograph that captures the most spectrally pure observations of the \halpha{} line to date. 
Launched in 2021, it has already collected an extensive database of observations, both synoptic and dedicated, that is now publicly available\footnote{\href{https://ssdc.nju.edu.cn}{Solar Science Data Center of Nanjing University}}.
In Figure~\ref{fig:chase_brightRim} we present two observations of a solar prominence taken with the \ac{HIS} instrument.
The entire disk passage of this prominence was recorded by the instrument, but we focus here on two specific times.
The first, presented in the left panel, is from 08 March 2023 when the prominence was close to the limb ($\approx$~60$^{\degree}$) but still projected against the solar disk as a filament.
At this time, the prominence is exhibiting the characteristic `bright-rim' phenomenon where the lowermost portion of the structure appears bright against both the prominence itself and the underlying solar chromosphere.
The second, presented in the right panel, is from 12 March 2023 when the same prominence is near disk-center ($\approx$~20$^{\degree}$).
At this time, there is no appearance of a `bright-rim' as typically described.
Inline with the discussion of \citet{Panasenco:2014}, however, by saturating the intensities of this observation we can reveal coherent brightenings that trace the outline of the prominence body, reminiscent of the more symmetric brightenings of Figure~\ref{fig:angularTransitions}.

Finally, it is clear that the appearance of the \Caii{~K} line in Figure~\ref{fig:spectralTransitions}, and especially Figure~\ref{fig:angularTransitions}, somewhat contradicts what we have come to anticipate from observations of solar prominences.
That is, the on-disk projection of a prominence (filament) in \Caii{~K} is almost exclusively an absorption signature yet appears in both figures with a positive constrast against the chromosphere \citep[cf.][]{Kuckein:2016, Chatterjee:2017, Chatzistergos:2023}.
For an identical model with an internal prominence pressure of 0.1~dyn~cm$^{-2}$, not explicitly shown here, we recover once more an absorption signature for the \Caii{~K} line.
It is clear that the `toy' model that we have adopted here, albeit employed as a proof of concept, is lacking a crucial dimension of realism.
One possibility is the idealised thermodynamic properties of the prominence, in particular the lack of an \ac{MHS} initial condition \citep[such as in][]{Heinzel:2001} or a self-consistent \ac{MHD} solution \citep[cf.][]{Jenkins:2021,Jenkins:2022}.
Another, as the latter works also highlight, is the assumed prominence topology of a monolithic versus threaded body, altering the opacity distribution along the \ac{LOS} and hence the radiative behaviour \citep[in line with the recent work of][]{Peat:2023}.
Furthermore, although the numerical resolution of the model was deliberately focused at the \ac{PCTR}, these authors have also shown that such a region can have a dimension of a few tens of km, far below the resolution of our model.
Each of these aspects will receive dedicated attention moving forward.

%
%

\section*{Acknowledgements}
RK and JMJ are supported by the ERC Advanced Grant PROMINENT and an FWO grant G0B4521N. This project has received funding from the European Research Council (ERC) under the European Union’s Horizon 2020 research and innovation programme (grant agreement No. 833251 PROMINENT ERC-ADG 2018). This research is further supported by Internal funds KU Leuven, project C14/19/089 TRACESpace. The computational resources and services used in this work were provided by the VSC (Flemish Supercomputer Center), funded by the Research Foundation Flanders (FWO) and the Flemish Government – department EWI. CMJO acknowledges support from the University of Glasgow’s College of Science and Engineering. CHASE mission is supported by China National Space Administration (CNSA). CL and YQ are supported by NSFC under grant 12333009 and CNSA project D050101.

\facilities{Vlaams Supercomputer Centrum, CHASE/HIS}


\software{Python \citep[][]{vanRossum:1995}, Lightweaver \citep[][]{Osborne:2021}, Numpy \citep[][]{Harris:2020}, Matplotlib \citep[][]{Hunter:2007}, yt-project \citep[][]{Turk:2011}}

\newpage
\bibliography{bibliography}{}
\bibliographystyle{aasjournal}



\end{document}

%% file: acronyms.tex
\usepackage{acro2}

\acsetup{
	single        = false,
	group-citation    = true,
	group-cite-cmd = \citealp,
	group-cite-connect = {; },
}

\DeclareAcronym{EVE}{
  short = EVE,
  long = \textit{Extreme Ultraviolet Variability Experiment},
  cite = {Woods:2012}}

\DeclareAcronym{AIA}{
  short = AIA,
  long = \textit{Atmospheric Imaging Assembly},
  cite = {Lemen:2012}}

\DeclareAcronym{HMI}{
  short = HMI,
  long = \textit{Helioseismic Magnetic Imager},
  cite = {Scherrer:2012}}
  
\DeclareAcronym{SDO}{
  short = SDO,
  long = \textit{Solar Dynamics Observatory},
  cite = {Pesnell:2012}}

\DeclareAcronym{STEREO}{
  short = STEREO,
  long = \textit{Solar Terrestrial Relations Observatory},
  cite = {Kaiser:2008}}

\DeclareAcronym{SECCHI}{
  short = SECCHI,
  long = \textit{Sun Earth Connection Coronal and Heliospheric Investigation},
  cite = {Howard:2008}}
  
\DeclareAcronym{SCIP}{
  short = SCIP,
  long = Sun Centered Imaging Package}
  
\DeclareAcronym{MGN}{
  short = MGN,
  long = multi-scale Gaussian normalisation,
  cite = {Morgan:2014}}

\DeclareAcronym{EUVI}{
  short = EUVI,
  long = EUVI}

\DeclareAcronym{CME}{
  short = CME,
  short-plural-form = CMEs,
  long = coronal mass ejection,
  long-plural-form = coronal mass ejections}
  
\DeclareAcronym{PIL}{
  short = PIL,
  short-plural-form = PILs,
  long = polarity inversion line,
  long-plural-form = polarity inversion lines}
  
\DeclareAcronym{LOS}{
  short = LOS,
  short-plural-form = LOS,
  long = line of sight,
  long-plural-form = lines of sight}
  
\DeclareAcronym{FOV}{
  short = FOV,
  short-plural-form = FOV,
  long = field of view,
  long-plural-form = fields of view}
  
\DeclareAcronym{EUV}{
  short = EUV,
  short-plural-form = EUV,
  long = extreme ultraviolet,
  long-plural-form = extreme ultraviolet}
  
\DeclareAcronym{GONG}{
  short = GONG,
  long = \textit{Global Oscillations Network Group}}
  
\DeclareAcronym{DST}{
  short = DST,
  long = \textit{Dunn Solar Telescope}}

\DeclareAcronym{IBIS}{
  short = IBIS,
  long = \textit{Interferometric Bidimensional Spectropolarimeter},
  cite = {Cavallini:2006,Reardon:2008}}

\DeclareAcronym{FIRS}{
  short = FIRS,
  long = \textit{Facility Infrared Spectropolarimeter},
  cite = {Jaeggli:2011}}
  
\DeclareAcronym{EIT}{
  short = EIT,
  long = \textit{Extreme Ultraviolet Imaging Telescope},
  cite = {Delaboudiniere:1995}}
 
\DeclareAcronym{ROSA}{
  short = ROSA,
  long = \textit{Rapid Oscillations in the Solar Atmosphere},
  cite = {Jess:2010}}
  
\DeclareAcronym{SPINOR}{
  short = SPINOR,
  long = \textit{Spectro-Polarimeter for Infrared and Optical Regions},
  cite = {Socas:2006}}  
 
\DeclareAcronym{SOHO}{
  short = SOHO,
  long = \textit{Solar and Heliospheric Observatory},
  cite = {Domingo:1995}}
 
\DeclareAcronym{MDI}{
  short = MDI,
  long = \textit{Michelson Doppler Imager},
  cite = {Scherrer:1995}}

\DeclareAcronym{NMSU}{
  short = NMSU,
  long = New Mexico State University}

\DeclareAcronym{NSO}{
  short = NSO,
  long = \textit{National Solar Observatory}}
  
\DeclareAcronym{TAC}{
  short = TAC,
  long = time allocation committee}
  
\DeclareAcronym{DEM}{
  short = DEM,
  short-plural-form = DEMs,
  long = Differential Emission Measure,
  long-plural-form = Differential Emission Measures}
  
\DeclareAcronym{PI}{
  short = PI,
  short-plural-form = PI,
  long = Principle Investigator,
  long-plural-form = Principle Investigators}  
  
\DeclareAcronym{IDL}{
  short = IDL,
  long = Interactive Data Language}
  
\DeclareAcronym{bb}{
  short = bb,
  long = broadband}  
  
\DeclareAcronym{nb}{
  short = nb,
  long = narrowband}    
  
\DeclareAcronym{IPM}{
  short = IPM,
  long = interplanetary medium}  
  
\DeclareAcronym{AO}{
  short = AO,
  long = adaptive optics,
  cite = {Rimmele:2004}} 
  
\DeclareAcronym{IR}{
  short = IR,
  long = infrared}  

\DeclareAcronym{halpha}{
  short = H-$\alpha$,
  long = Hydrogen-$\alpha$}

\DeclareAcronym{FPI}{
  short = FPI,
  long = Fabry-P\'erot}
  
\DeclareAcronym{DWDM}{
  short = DWDM,
  long = dense wavelength division multiplexing}  

\DeclareAcronym{CCD}{
  short = CCD,
  long = charge-coupled device}   
  
\DeclareAcronym{HI}{
  short = HI,
  long = Heliospheric Investigation}   
  
\DeclareAcronym{ROI}{
  short = ROI,
  long = region-of-interest}    
  
\DeclareAcronym{POV}{
  short = POV,
  long = point-of-view}  
  
\DeclareAcronym{MHS}{
  short = MHS,
  long = magnetohydrostatic}
  
\DeclareAcronym{MHD}{
  short = MHD,
  long = magnetohydrodynamic}  
  
\DeclareAcronym{RMHD}{
  short = RMHD,
  long = radiative magnetohydrodynamic}  
  
\DeclareAcronym{DOT}{
  short = DOT,
  long = \textit{Dutch Open Telescope},
  cite = {Rutten:1997}}  
  
\DeclareAcronym{BBSO}{
  short = BBSO,
  long = \textit{Big Bear Solar Observatory}}
  
\DeclareAcronym{NST}{
  short = NST,
  long = \textit{New Solar Telescope},
  cite = {Goode:2012}}

\DeclareAcronym{SOT}{
  short = SOT,
  long = \textit{Solar Optical Telescope},
  cite = {Tsuneta:2008}}
  
\DeclareAcronym{hinode}{
  short = Hinode,
  long = Hinode,
  cite = {Kosugi:2007}}
  
\DeclareAcronym{GST}{
  short = GST,
  long = \textit{Goode Solar Telescope}}  
  
\DeclareAcronym{SST}{
  short = SST,
  long = \textit{Swedish 1-m Solar Telescope},
  cite = {Scharmer:2002}}  
  
\DeclareAcronym{TRACE}{
  short = TRACE,
  long = \textit{Transition Region and Coronal Explorer},
  cite = {Handy:1999}}  
  
\DeclareAcronym{PCTR}{
  short = PCTR,
  long = \textit{prominence-corona transition region}}
  
\DeclareAcronym{DKIST}{
  short = DKIST,
  long = \textit{Daniel K. Inouye Solar Telescope},
  cite = {Rimmele:2020}}
  
\DeclareAcronym{RTI}{
  short = RTI,
  long = Rayleigh-Taylor instability}
  
\DeclareAcronym{SSW}{
  short = SSW,
  long = SolarSoftWare,
  cite = {Freeland:1998}}    
  
\DeclareAcronym{LTE}{
  short = LTE,
  long = local thermodynamic equilibrium}
  
\DeclareAcronym{NLTE}{
  short = NLTE,
  long = outside local thermodynamic equilibrium}
  
\DeclareAcronym{FTS}{
  short = FTS,
  long = Fourier Transform Spectrometer,
  cite = {Kurucz:1984}}
  
\DeclareAcronym{RTE}{
  short = RTE,
  long = radiative transfer equation}
  
\DeclareAcronym{CRD}{
  short = CRD,
  long = complete frequency redistribution}
  
\DeclareAcronym{PRD}{
  short = PRD,
  long = partial frequency redistribution}
  
\DeclareAcronym{BCM}{
	short = BCM,
	long = Beckers' cloud model,
	cite={Beckers:1964}}
	
\DeclareAcronym{HSRA}{
	short = HSRA,
	long = Harvard Smithsonian Reference Atmosphere,
	cite={Gingerich:1971}}
	
\DeclareAcronym{NICOLE}{
	short = NICOLE,
	long = Non-LTE Inversion COde using the Lorien Engine,
	cite={Socasnavarro:2015}}
	
\DeclareAcronym{SIR}{
	short = SIR,
	long = Stokes Inversion based on Response functions,
	cite={Ruizcobo:1992}}
	
\DeclareAcronym{HAZEL}{
	short = HAZEL,
	long = Hanle and Zeeman Light,
	cite={AsensioRamos:2008}}
	
\DeclareAcronym{BPSS}{
	short = BPSS,
	long = bald patch separatrix surface,
	cite={Bungey:1996}}
	
\DeclareAcronym{EIS}{
	short = EIS,
	long = \textit{EUV Imaging Spectrometer},
	cite={Culhane:2007}}

\DeclareAcronym{FWHM}{
  short = FWHM,
  long = full width at half maximum}

\DeclareAcronym{AMRVAC}{
	short = MPI-AMRVAC,
	long = \textit{Adaptive Mesh Refinement Versatile Advection Code},
	cite = {Keppens:2012,Porth:2014,Xia:2018,Keppens:2020}}

\DeclareAcronym{AMR}{
	short = AMR,
	long = adaptive mesh refinement}

\DeclareAcronym{CCI}{
	short = CCI,
	long = Convective Continuum Instability}

\DeclareAcronym{BV}{
	short = BV,
	long = Brunt-V\"ais\"al\"a}

\DeclareAcronym{TVDLF}{
	short = TVDLF,
	long = Total Variation Diminishing Lax-Friedrich}

\DeclareAcronym{TI}{
	short = TI,
	long = Thermal Instability}

\DeclareAcronym{TNE}{
	short = TNE,
	long = thermal non-equilibrium}
	
\DeclareAcronym{MALI}{
    short = MALI,
    long = Multilevel Accelerated Lambda Iteration}
    
\DeclareAcronym{yt}{
    short = yt,
    long = yt-project,
    cite = {Turk:2011}}
    
\DeclareAcronym{GL}{
    short = GL,
    long = Gauss-Legendre}
    
\DeclareAcronym{CM}{
    short = CM,
    long = classically mounted}
    
\DeclareAcronym{MULTI3D}{
    short = MULTI3D,
    long = \textit{multi-level non-LTE 3D},
    cite = {Leenaarts:2009}}

\DeclareAcronym{EB}{
    short = EB,
    long = \textit{Eddington-Barbier}}
    
\DeclareAcronym{KDE}{
    short = KDE,
    long = kernel density estimate}
    
\DeclareAcronym{NLFFF}{
    short = NLFFF,
    long = non-linear force-free field}

\DeclareAcronym{EOS}{
    short = EOS,
    long = equation-of-state
}

\DeclareAcronym{SPICE}{
    short = SPICE,
    long = \textit{SPectral Imaging of the Coronal Environment},
    cite = {Spice:2020}
}
\DeclareAcronym{SolO}{
	short = SolO,
	long = \textit{Solar Orbiter},
	cite={Muller:2020}
}
\DeclareAcronym{VTF}{
	short = VTF,
	long = \textit{Visible Tunable Filter},
	cite={Schmidt:2014}
}
\DeclareAcronym{CHASE}{
  short = CHASE,
  long = \textit{Chinese H$\alpha$ Solar Explorer},
  cite = {LiChuan:2022}
}
\DeclareAcronym{HIS}{
  short = HIS,
  long = \textit{H$\alpha$ Imaging Spectrograph},
  cite = {LiuQiang:2022}
}